\pdfoutput=1
\documentclass{my-paper}

\title{Token Composition: A Graph Based on \glstext*{evm} Logs\thanks{This is a
preprint of a published article~\citep{harrigan-et-al-24}.}}

\author{Martin~Harrigan\thanks{\email{martin.harrigan@setu.ie},
\orcid{0000-0002-6069-7001}}}

\author{Thomas~Lloyd\thanks{\email{tomlloyd1992@gmail.com}}}

\author{Daire~\'O~Broin\thanks{\email{daire.obroin@setu.ie},
\orcid{0000-0002-2886-5546}}}

\affil{\gls*{setu}, \acrlong*{roi}}

\date{}

\begin{document}

\maketitle

\begin{abstract}
  Tokens have proliferated across blockchains in terms of number, market
capitalisation, and utility.  Some tokens are tokenised versions of
existing tokens, known variously as wrapped tokens, fractional tokens
or shares.  The repeated application of this process creates tokens
with arbitrarily many layers of composition.  We perform an empirical
analysis of token composition on the Ethereum blockchain.  We
introduce a graph that represents the tokenisation of tokens by other
tokens, and we show that the graph contains non-trivial topological
structure.  We relate properties of the graph, for example, connected
components and cyclic structure, to the tokenisation process.  For
example, we identify the longest directed path and its corresponding
sequence of tokens, and we visualise the connected components relating
to a stablecoin and a non-fungible token protocol.  Our goal is to explore and
visualise what has been built with tokens, rather than propose new
constructions.

\end{abstract}

\section{Introduction}\label{sec:introduction}

\begin{figure}
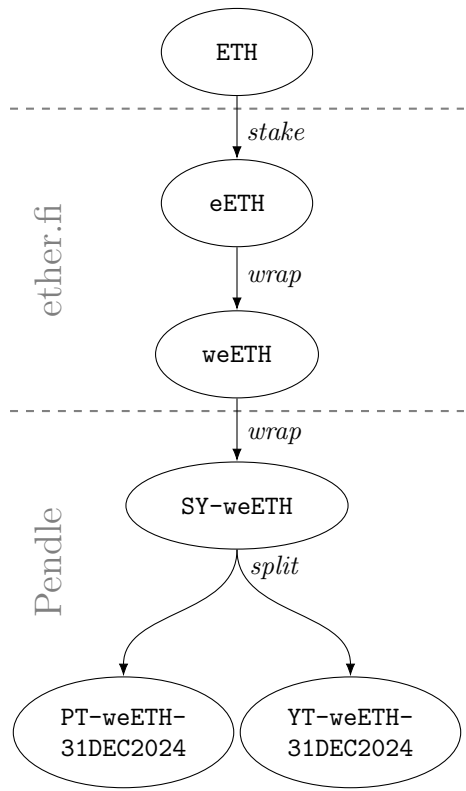

  \centering
  \myincludegraphics[width=0.5\linewidth]{tikz/defi/pt-yt-weeth}
  \caption{Tokens can have many layers of composition.  For example,
    one can: \emph{stake} \unit{\eth} for \unit{\eeth} to earn yield;
    \emph{wrap} \unit{\eeth} for \unit{\weeth} to collect the yield;
    \emph{wrap} \unit{\weeth} for \unit{\syweeth} to standardise the
    yield collection mechanism; and \emph{split} \unit{\syweeth} into
    \unit{\ptweeth} and \unit{\ytweeth} to separate the principal from
  the yield up to a maturity.}\label{fig:pt-yt-weeth}
\end{figure}

The number of tokens on blockchains has grown rapidly: Ethereum
alone hosts hundreds of thousands of ERC-20 tokens.  Many are
simple, in that they are not composed of other tokens.  Others are
composed: a liquidity pool token, for example, represents a share of
a collection of other tokens, and \citet{dex-screener-xx}, a popular
liquidity pool tracker, lists over one hundred thousand of them.
Tokens are also being composed in more elaborate ways:
\unit{\ptweeth\text-25APR2024}~(\texttt{0xb18c87}) is a token issued
by Pendle Finance~\citep{nguyen-vuong-22} on the Arbitrum blockchain
that transitively depends on many other tokens (see
\cref{fig:pt-yt-weeth}).  At the time of writing, this token has a
market capitalisation of over \dUSD[prefix=billion~]{1}\@.
Examining token composition matters for both technical and financial
risk: if an investor purchases \unit{\ptweeth\text-25APR2024}, what
tokens does the investment depend upon?  In this paper, we present a
method for examining token composition at both the macro- and
micro-level and apply it to the Ethereum blockchain.

Our method extracts \emph{meta-events} from \gls*{evm} logs.
Contracts emit low-level events; the ERC-20 standard, for example,
specifies a \texttt{Transfer} event whenever a token is
transferred~\citep{vogelsteller-buterin-15}.  A meta-event is a
heuristic-defined sequence of one or more low-level events emitted by
a single transaction.  A meta-event could signify a token being
tokenised by another token, that is, a deposit of an underlying token
with a contract and the minting of a new share, or the burning of a
share and the withdrawal of an underlying token from a contract.
This meta-event, which we call a \emph{tokenising meta-event}, can be
identified from multiple \texttt{Transfer} events within a single
transaction.  The tokenising meta-events can be represented as a
\emph{token graph}: each vertex represents a token and each directed
edge represents the token corresponding to the source vertex being
tokenised by the token corresponding to the target vertex.  We apply
various forms of graph analysis to the token graph and visualise the
structure of token compositions.

This paper is organised as follows.  In \cref{sec:related-work} we
review related work.  In \cref{sec:token-composition} we introduce
token composition, the token graph and our data sources.  We present
our analysis in \cref{sec:analysis}.  Finally, we conclude in
\cref{sec:conclusions}.

\section{Related Work}\label{sec:related-work}

We categorise related work into four areas: empirical analysis of
smart contract composition and code reuse, the automatic detection of
tokens, graph analysis of blockchains and token systems, and wrapped
tokens.

Software composition is a hard problem~\citep{garlan-et-al-94}.  Smart
contracts sidestep the low-level problems of interoperability by using
a shared execution environment (that is, a virtual machine) and
\textit{de facto} standards (for example, ERCs), and the high-level problems
of architectural mismatch by taking a bottom-up approach to
composition.  \citet{he-et-al-20} perform a large-scale analysis of
\num{10} million Ethereum smart contracts deployed between July 2015
and December 2018.  They show that less than \qty{1}{\percent} of the
contracts are distinct, and more than \qty{63}{\percent} of those are
similar to at least one other contract.  The results have been
replicated (see, for example,
\citet{kondo-et-al-20,chen-et-al-21,khan-et-al-22}).

\citet{frowis-et-al-19} use symbolic execution to automatically detect
smart contracts on the Ethereum blockchain that implement token
functionality.  \citet{di-angelo-salzer-23} reconstruct contract
interfaces and events from \glstext*{evm} bytecode.  They use transaction data
to identify token contracts that comply with ERC standards and token
contracts that do not.  \citet{oliveira-et-al-18} propose a taxonomy
for classifying tokens and a decision tree to guide the
token design process.

\citet{kitzler-et-al-21} analyse activity relating to
\emph{decentralised finance} on the Ethereum blockchain.  They
construct and topologically analyse two graphs: the \emph{contract
account graph} where the vertices are contract accounts and the edges
are transactions between those accounts, and the \emph{protocol graph}
where the vertices are protocols and the edges are transactions
between those protocols.  They show that community finding algorithms
identify communities in the contract account graph, but the
communities do not correspond to protocols.  There are several network
analyses of ERC-20 tokens on the Ethereum blockchain that quantify
their age, economic value, and activity volume (see, for example,
\citet{somin-et-al-18,victor-luders-19}).

\citet{caldarelli-21} describes wrapped tokens and their ability to
represent real-world assets and to bridge tokens across blockchains.
The \unit{\wbtc} whitepaper~\citep{kyber-et-al-xx} sets out a general
framework for tokenising assets on a blockchain.
\citet{santoro-et-al-22} propose a standard interface for
\emph{vaults} for ERC-20 tokens.  A vault can store a single
\emph{asset} or underlying token.  Users can \emph{deposit} or
\emph{withdraw} the asset.  In return, they receive \emph{shares} in
the form of another ERC-20 token.  \citet{lloyd-et-al-23} analyse the
emergent outcomes of \emph{vote-escrowed tokens} (veTokens) where a
token is locked for a fixed period in exchange for voting rights.

We use common terminology from graph theory throughout the paper.
See \citet{diestel-17} for definitions.

\section{Token Composition}\label{sec:token-composition}

Tokens are central to blockchain-based protocols and
applications~\citep{voshmgir-20}.  Token composition is a technique
where one or more tokens can be combined to create new tokens.  The
entitlements of the input tokens are appropriated by the new token.
We extract data from \glstext*{evm} logs (\cref{sec:token-comp-evm-logs}) and use
it to construct a token graph (\cref{sec:token-comp-token-graph}), a
representation of the tokenisation process.  In \cref{sec:analysis} we
analyse the graph and relate it to the tokenisation process.

\subsection{\glstext*{evm} Logs, Events and Meta-Events}\label{sec:token-comp-evm-logs}

\glstext*{evm} logs record specific occurrences or outputs generated during the
execution of contract code.  They enable off-chain applications to
react to on-chain events.  A popular event is the \texttt{Transfer}
event emitted by ERC-20 tokens~\citep{vogelsteller-buterin-15}:

\begin{lstlisting}[language=solidity,caption={The ERC-20 \texttt{Transfer} event specifies three
      parameters: \texttt{from}, \texttt{to}, and \texttt{value}.}]
  event Transfer(address indexed from, address indexed to, uint256 value);
\end{lstlisting}

The event has two special cases.  If the \texttt{from} address is the
zero address (\texttt{0x0}) then the contract mints new tokens.  If
the \texttt{to} address is the zero address then it burns existing
tokens.

A single transaction can emit multiple events.  We define a
\emph{meta-event} to be a sequence of events that match some pattern
and are emitted by a single transaction.  We define a \emph{tokenising
meta-event} to be a meta-event whose pattern is one of two pairs of
\texttt{Transfer} events.  The first pair, \emph{deposit \& mint},
consists of a transfer of tokens to the contract and a new token
being minted.  The second pair, \emph{withdraw \& burn}, consists of
a token being burned and a transfer of an existing token from the
contract.  We want to identify instances where a token is tokenised
by another.  In the terminology of ERC-4626~\citep{santoro-et-al-22},
a tokenising meta-event corresponds to either a \texttt{Deposit}
event or a \texttt{Withdraw} event.  However, our tokenising
meta-event does not require the contract to follow ERC-4626 and emit
those precise events.

\afterpage{%
\begin{landscape}
  \begin{table*}
    \centering
    \caption{Each tokenising meta-event contains the address of the
      source token, the address of the target token, one of two possible
      pairs of actions (deposit \& mint or withdraw \& burn), the amount
      of the source token that was deposited or withdrawn, the amount of
      the target token that was minted or burned, and a transaction
      hash.  The table includes four sample entries from the full set of
      \num{4032033} tokenising meta-events: the earliest and latest
      tokenising meta-events that have deposit \& mint and withdraw \&
    burn actions.}\label{tab:meta-events}
    \begin{tabular}{cccrrc}
      \toprule
      Source Token &
      Target Token &
      Actions &
      Source Amount &
      Target Amount &
      Tx Hash\\
      \midrule

      \unit{\arc}~(\texttt{0xac709f}) & \unit{\swt}~(\texttt{0xb12a3c})
      & deposit \& mint & \emph{dust} & \emph{dust} &
      \texttt{0x549a12}\\

      \unit{\dgz}~(\texttt{0x84178d}) &
      \unit{\predgz}~(\texttt{0x18aa6e}) & withdraw \& burn & \num{1371}
      & \num{150} & \texttt{0x2da232}\\

      \unit{\bone}~(\texttt{0x981303}) &
      \unit{\tbone}~(\texttt{0xf7a038}) & withdraw \& burn & \num{5183}
      & \num{5160} & \texttt{0x5dbe32}\\

      \unit{\weth}~(\texttt{0xc02aaa}) &
      \unit{\aweth}~(\texttt{0x030ba8}) & deposit \& mint & \num{25} &
      \num{25} & \texttt{0xb4281a}\\

      \bottomrule
    \end{tabular}
  \end{table*}
\end{landscape}%
}

We extracted all \texttt{Transfer} events from Ethereum mainnet for
block heights \numrange{0}{16685101} (February 2023) inclusive using
Geth's \texttt{eth\_getLogs} \gls*{rpc} method~\citep{go-ethereum-xx}.  From
the \texttt{Transfer} events, we identified \num{4032033} tokenising
meta-events using the pattern described above.  \Cref{tab:meta-events}
shows a sample of the data.  The first row is the earliest deposit \&
mint and the second is the earliest withdraw \& burn; the third is
the latest withdraw \& burn and the fourth is the latest deposit \&
mint.

For example, the first row indicates a transaction that deposited a
dust amount of \unit{\arc}~\citep{arcade-city-xx} to a contract in
exchange for newly minted \unit{\swt}~\citep{swarm-city-xx} in
January 2017.  The third row indicates a transaction that withdrew
\qty{5183}{\bone}~\citep{shiba-inu-xx} from a contract in exchange for
burning \qty{5160}{\tbone} in February 2023.  We do not consider the
individual utility or value of the tokens, only whether Token~$X$ was
deposited with a contract to mint Token~$Y$, and/or Token~$Y$ was
burned by a contract to withdraw Token~$X$.

We filter the tokenising meta-events to retain only those involving a
pair of tokens, Token~$X$ and Token~$Y$, that
admit both directions: at least one deposit of Token~$X$ to
mint Token~$Y$, and at least one burn of
Token~$Y$ to withdraw Token~$X$.  In other words,
the \textquote{and/or} conjunction in the previous paragraph is
replaced by \textquote[][.]{and} This excludes two cases.
\emph{One-way token upgrades} are pairs where Token~$X$ can
be deposited to mint Token~$Y$, but Token~$X$
cannot be withdrawn from the contract.  \emph{One-way token burns}
are pairs where Token~$Y$ can be burned to withdraw
Token~$X$, but Token~$X$ cannot be deposited to
mint Token~$Y$.  Of the \num{4032033} tokenising meta-events,
\num{3461723} meet the additional requirement.  We refer to the
\emph{unfiltered} and \emph{filtered tokenising meta-events} in
\cref{sec:token-comp-token-graph}.

We also incorporate market capitalisation data from
\citet{gecko-labs-xx} and liquidity pool data from
\citet{dex-screener-xx}.  CoinGecko aggregates token data including
market price, exchange volume, and market capitalisation.  \glstext*{dex}
Screener stores, parses, and analyses blockchain data to produce a
token screener with charts.

\subsection{The Token Graph}\label{sec:token-comp-token-graph}

\begin{figure}
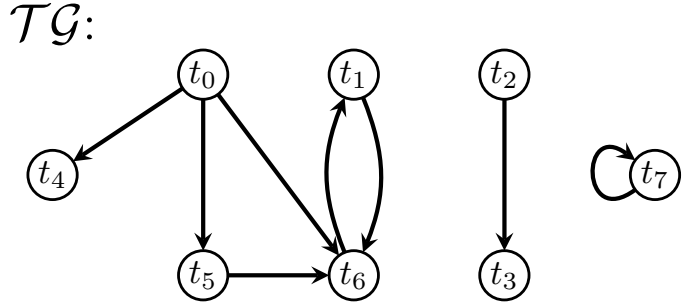

  \centering
  \myincludegraphics[width=0.75\linewidth]{tikz/defi/token-graph}
  \caption{A token graph, $\mathit{TG}$, with eight vertices, $t_0,
    t_1, \ldots, t_7$, and eight edges.  Each edge corresponds to a
    token tokenising another token.  For example, the token
    represented by $t_5$ tokenises the token represented by
  $t_0$.}\label{fig:token-graph}
\end{figure}

We can construct a directed graph from the tokenising meta-events as
follows.  Each vertex corresponds to a token.  Each directed edge from
a source vertex to a target vertex corresponds to a set of tokenising
meta-events that deposits the source token and mints the target token,
and/or withdraws the source token and burns the target token.  In the
terminology of ERC-4626, the source token is the \emph{asset} and the
target token is the \emph{share}.  \Cref{fig:token-graph} shows an
example token graph.

In the unfiltered case, the directed graph has \num{23687} vertices
(distinct tokens) and \num{23549} edges; each edge represents a pair
of tokens where the second tokenises the first with either a deposit
\& mint or a withdraw \& burn action.  Filtering to retain only edges
where both actions are observed yields \num{7536} edges connecting
\num{8424} of the original vertices.  We refer to the
\emph{unfiltered} and \emph{filtered token graphs} in the remainder
of the paper.

The filtered token graph is a higher-confidence subset of the
unfiltered token graph, not a complete representation of token
composition on the Ethereum blockchain.  Requiring both a
tokenisation operation and its inverse reduces false positives, but
it also removes legitimate compositions whose inverses do not appear
in the data.  The unfiltered token graph offers greater coverage; the
filtered token graph offers greater confidence.

\subsection{Data Limitations}\label{sec:token-comp-limitations}

Our input data, namely, \glstext*{evm} logs, CoinGecko market data, and \glstext*{dex}
Screener liquidity pool data, have limitations.  First, \glstext*{evm} logs are
unauthenticated: a contract can emit an event of its choosing.  There
is no guarantee that, say, an ERC-20 \texttt{Transfer} event
accurately reflects an actual transfer~\citep{guidi-michienzi-22}.
Furthermore, the first special case highlighted in
\cref{sec:token-comp-evm-logs} (mint) is recommended by
ERC-20\footnote{\textquote[\citep{vogelsteller-buterin-15}]{A token
    contract which creates new tokens SHOULD trigger a Transfer event
    with the \texttt{\_from} address set to \texttt{0x0} when tokens are
created.}} but the second (burn) is not.  However, \glstext*{evm} logs for the
Ethereum blockchain are generally accurate and malicious contracts can
be easily excluded.  For an aggregated analysis, such as ours, we
believe the impact of these limitations is minimal.

Second, the data from CoinGecko and \glstext*{dex} Screener are snapshots that
were gathered in April 2024 whereas the \glstext*{evm} logs have a temporal
component.  It is possible that a token had an entry on CoinGecko in
the past, but, at the time the data was gathered, the entry no longer
existed.  It is also possible that a token was tracked by \glstext*{dex} Screener
in the past, but, at the time the data was gathered, it was no longer
being tracked.  It is also possible that the token coverage of
CoinGecko or \glstext*{dex} Screener is incomplete or inaccurate.  However, as a
high-level measure of token popularity, we believe the impact of this
potential mismatch is minimal.

Third, tokenising meta-events are a heuristic for identifying
instances where one token is tokenised by another.  False positives
create edges in the token graph where the token corresponding to the
source is not tokenised by the token corresponding to the target;
false negatives are pairs of tokens where one is tokenised by the
other but there is no corresponding edge in the token graph.  We
discuss both cases in \cref{sec:analysis} and \cref{sec:conclusions}.

\section{Analysis}\label{sec:analysis}

In this section we examine the macro-topological structure of the
unfiltered and filtered token graphs including degree distributions,
connected component structure, and cyclic structure.  We also examine
the micro-topological structure of some individual tokens.  We
visualise the composition of tokens and identify their direct and
transitive dependencies.

\begin{table}
  \centering
  \caption{Each edge in a token graph represents a set of tokenising
    meta-events.  The top five most significant edges in terms of the
    number of tokenising meta-events they contain are shown.  The
    results are the same for both the unfiltered and filtered token
    graphs, that is, the same five edges are present in
  both.}\label{tab:most-significant-edges}
  \begin{tabular}{ccr}
    \toprule

    Source Token & Target Token &
    \begin{tabular}{@{}c@{}}\#
      Tokenising\\Meta-Events
    \end{tabular}\\

    \midrule

    \unit{\shib}~(\texttt{0x95ad61}) &
    \unit{\xshib}~(\texttt{0xb4a812}) & \num{402186}\\

    \unit{\bone}~(\texttt{0x981303}) &
    \unit{\tbone}~(\texttt{0xf7a038}) & \num{203734}\\

    \unit{\sushi}~(\texttt{0x6b3595}) &
    \unit{\xsushi}~(\texttt{0x879824}) & \num{120221}\\

    \unit{\leash}~(\texttt{0x27c70c}) &
    \unit{\xleash}~(\texttt{0xa57d31}) & \num{75180}\\

    \unit{\usdc}~(\texttt{0xa0b869}) &
    \unit{\ausdc}~(\texttt{0xbcca60}) & \num{69373}\\

    \bottomrule
  \end{tabular}
\end{table}

Each edge in a token graph represents a set of tokenising
meta-events.  Before examining the structure of the graphs, we
identify the most significant edges in terms of the number of
tokenising meta-events they contain.
\Cref{tab:most-significant-edges} shows the results for both the
unfiltered and filtered token graphs.  Three of the five edges
represent the staking of memecoins
(\unit{\shib}~$\rightarrow$~\unit{\xshib},
  \unit{\bone}~$\rightarrow$~\unit{\tbone}, and
\unit{\leash}~$\rightarrow$~\unit{\xleash}), one represents the
staking of a governance token for a decentralised exchange
(\unit{\sushi}~$\rightarrow$~\unit{\xsushi}), and one represents the
supply of a stablecoin to a decentralised lending market
(\unit{\usdc}~$\rightarrow$~\unit{\ausdc}).  We could also measure the significance of an edge based on, say, the
volume of tokens transacted or the present USD value of the locked
tokens.

\subsection{Degree Distributions}\label{sec:analysis-degree-distribution}

\begin{figure}
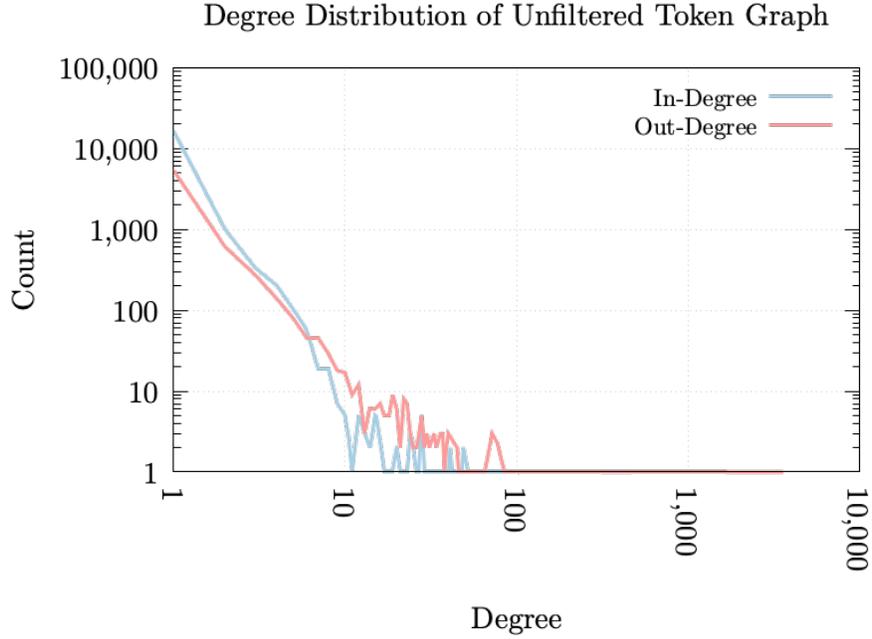

  \centering
  \myincludegraphics[width=\linewidth]{gnuplot/token-graphs/unfiltered-token-graph-degrees}
  \caption{The in- and out-degree distributions of the unfiltered
    token graph show an inverse relationship between the degree of a
    vertex and the number of vertices with that degree.  Few vertices
    have high degree and many have low
  degree.}\label{fig:unfiltered-token-graph-degrees}
\end{figure}

\begin{figure}
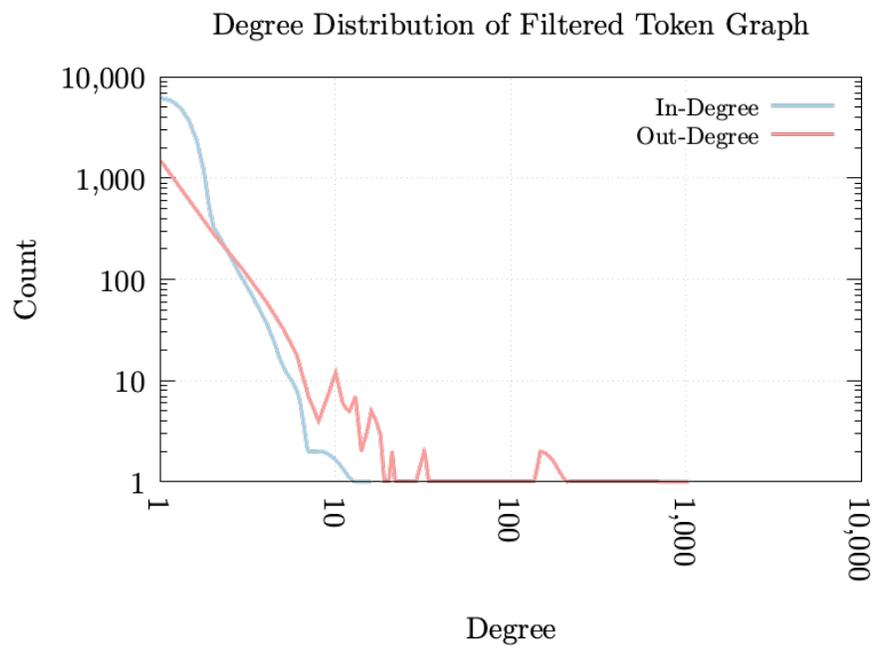

  \centering
  \myincludegraphics[width=\linewidth]{gnuplot/token-graphs/filtered-token-graph-degrees}
  \caption{The in- and out-degree distributions of the filtered token
    graph show a similar inverse relationship as in
  \cref{fig:unfiltered-token-graph-degrees}.}\label{fig:filtered-token-graph-degrees}
\end{figure}

\begin{table}
  \centering
  \caption{The top five vertices (tokens) in the unfiltered and
    filtered token graphs by in-degree and out-degree.  The truncated
    address for \unit{\chi} is genuinely all zeros: the full contract
    address is
  \texttt{0x0000000000004946c0e9f43f4dee607b0ef1fa1c}.}\label{tab:highest-in-out-degrees}
  \begin{tabular}{ccrcr}
    \toprule

    & \multicolumn{2}{c}{Top Five by In-Degree} &
    \multicolumn{2}{c}{Top Five by Out-Degree}\\

    \cmidrule(lr){2-3}\cmidrule(lr){4-5}

    & Token & Deg. & Token & Deg.\\

    \midrule

    \multirow{5}{0pt}{\rotatebox{90}{Unfiltered}}

    & \unit{\chi}~(\texttt{0x000000}) & \num{471}
    & \unit{\usdc}~(\texttt{0xa0b869}) & \num{3587}\\

    & \unit{\usdp}~(\texttt{0x145668}) & \num{117} &
    \unit{\dai}~(\texttt{0x6b1754}) & \num{1923}\\

    & \unit{\ausdc}~(\texttt{0xbcca60}) & \num{84} &
    \unit{\usdt}~(\texttt{0xdac17f}) & \num{1175}\\

    & \unit{\aweth}~(\texttt{0x030ba8}) & \num{63} &
    \unit{\weth}~(\texttt{0xc02aaa}) & \num{951}\\

    & \unit{\adai}~(\texttt{0x028171}) & \num{54} &
    \unit{\susd}~(\texttt{0x57ab1e}) & \num{548}\\

    \midrule

    \multirow{5}{0pt}{\rotatebox{90}{Filtered}}

    & \unit{\xdptwo}~(\texttt{0xe68c1d}) & \num{16} &
    \unit{\usdc}~(\texttt{0xa0b869}) & \num{1037}\\

    & \unit{\xdpone}~(\texttt{0x134fc6}) & \num{15} &
    \unit{\dai}~(\texttt{0x6b1754}) & \num{752}\\

    & \unit{\cyusd}~(\texttt{0x1d0914}) & \num{14} &
    \unit{\usdt}~(\texttt{0xdac17f}) & \num{396}\\

    & \unit{\idol}~(\texttt{0x7591a3}) & \num{13} &
    \unit{\weth}~(\texttt{0xc02aaa}) & \num{281}\\

    & \unit{\ageur}~(\texttt{0x1a7e4e}) & \num{8} &
    \unit{\wbtc}~(\texttt{0x2260fa}) & \num{211}\\

    \bottomrule
  \end{tabular}
\end{table}

The in- and out-degree distributions of the unfiltered and filtered
token graphs show an inverse relationship between the degree of a
vertex and the number of vertices with that degree (see
\cref{fig:unfiltered-token-graph-degrees,fig:filtered-token-graph-degrees}).
\Cref{tab:highest-in-out-degrees} shows the top five vertices in the
unfiltered and filtered token graphs by in-degree and out-degree.  The
out-degree entries are easy to explain: they are tokens that are
deposited with contracts to mint many other types of tokens.
They include stablecoins (\unit{\usdc}, \unit{\dai}, \unit{\usdt},
and \unit{\susd}), wrapped \unit{\eth} (\unit{\weth}), and wrapped
bitcoin (\unit{\wbtc}).  \Cref{fig:weth-usdc-composition} illustrates
two of these: \unit{\weth} is tokenised by \unit{\aweth},
\unit{\steth}, and \unit{\peth}, with \unit{\steth} itself tokenised
by \unit{\wsteth}; \unit{\usdc} is tokenised by \unit{\ausdc} and
\unit{\cusdc}\@.  Each edge in the figure corresponds to an edge in
the token graph, and the high out-degree of these base tokens
reflects the number of protocols that tokenise them.

\begin{figure}
  \centering
  \myincludegraphics[width=0.75\linewidth]{tikz/defi/weth-usdc-composition}
  \caption{Token composition for \unit{\weth} and \unit{\usdc}\@.
    \unit{\weth} is tokenised by \unit{\aweth}, \unit{\steth}, and
    \unit{\peth}; \unit{\steth} is tokenised by \unit{\wsteth}\@.
    \unit{\usdc} is tokenised by \unit{\ausdc} and \unit{\cusdc}\@.
    Each edge corresponds to an edge in the token
  graph.}\label{fig:weth-usdc-composition}
\end{figure}

The in-degree entries are more complex and have
multiple explanations.  For example, \unit{\chi}~\citep{1inch-20} is a
gas token created by 1inch, a decentralised exchange aggregator, that
is burned to obtain a reduction in transaction fees; in some
transactions the burning of \unit{\chi} is combined with the
withdrawal of another token.  This is a false positive generated by
our heuristic since \unit{\chi} does not tokenise a token.  The
remaining in-degree entries in the unfiltered category are due to
token swaps performed during a deposit.  For example,
\unit{\adai}~\citep{aave-xx} is a yield-bearing token issued by Aave,
a decentralised lending market, in exchange for the stablecoin
\unit{\dai}\@.  However, in some transactions, other tokens are
supplied and swapped to \unit{\dai}\@.  These are also false positives
since only \unit{\dai} is tokenised by \unit{\adai}\@.  In the
filtered category, the in-degree entries are more reliable.  For
example, the \unit{\idol} token is minted when a user deposits various
forms of the \unit{\sbt} token (for example, \unit{\sbt\relax~09180200},
\unit{\sbt\relax~09250200}, etc.) according to the Lien
Protocol~\citep{lien-20}.  Similarly, the \unit{\ageur} token is a
stablecoin issued by the Angle Protocol~\citep{angle-xx} that accepts
a variety of tokens as collateral.

\begin{figure}
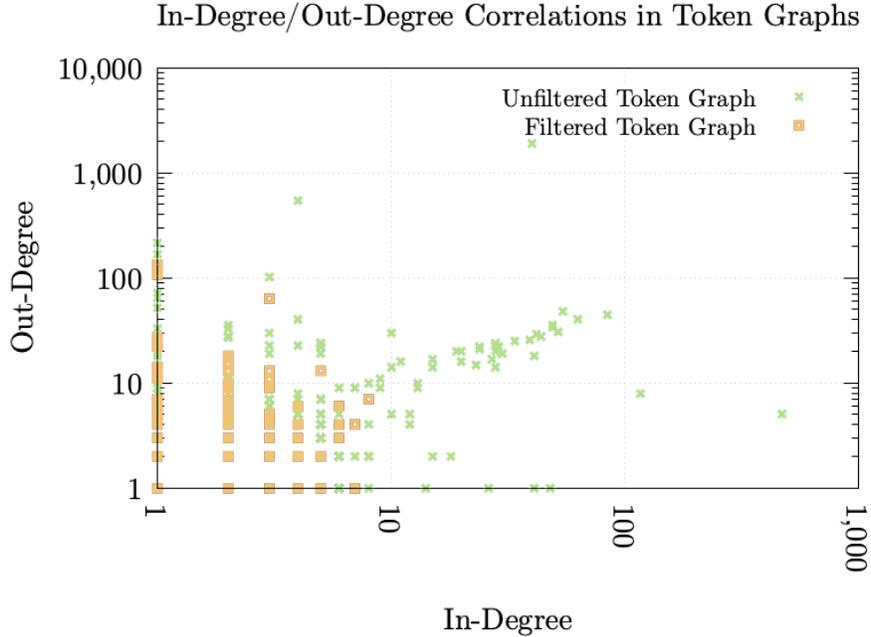

  \centering
  \myincludegraphics[width=\linewidth]{gnuplot/token-graphs/token-graph-in-out-degrees}
  \caption{In the unfiltered and filtered token graphs, we can
    identify vertices with both high in-degree and high
  out-degree.}\label{fig:token-graph-in-out-degrees}
\end{figure}

\Cref{fig:token-graph-in-out-degrees} plots in-degrees against
out-degrees in the unfiltered and filtered token graphs.  The tokens
whose corresponding vertices have both high in-degree and high
out-degree are tokens that tokenise many other tokens, and are
themselves tokenised by many other tokens.  Examples include
\unit{\musd}, a stablecoin issued by the mStable
protocol~\citep{mstable-xx} and the \unit{\ageur} token.  Stablecoins
can be minted from various forms of collateral and themselves used as
collateral to mint other tokens, which accounts for both their high
in- and out-degrees.

\subsection{Connected Component
Structure}\label{sec:analysis-component-structure}

The example token graph in \cref{fig:token-graph} contains three
weakly connected components: $\{t_0, t_1, t_4, t_5, t_6\}$, $\{t_2,
t_3\}$, and $\{t_7\}$.  The unfiltered graph has \num{4082} weakly
connected components.  A giant component contains \num{13794} vertices
(\qty{\sim58}{\percent}) and \num{17711} edges (\qty{\sim75}{\percent}).
There are \num{3336} components (\qty{\sim82}{\percent} of the total)
with only two connected vertices.  These vertices correspond to pairs
of tokens where at least one tokenises the other, but neither
tokenises, or is tokenised by, a third.  All tokens
represented by the vertices in these components have zero market
capitalisation according to CoinGecko, and none are traded in any
liquidity pool according to \glstext*{dex} Screener.  Their lack of popularity
reflects their isolation in the token graph.

\begin{figure}
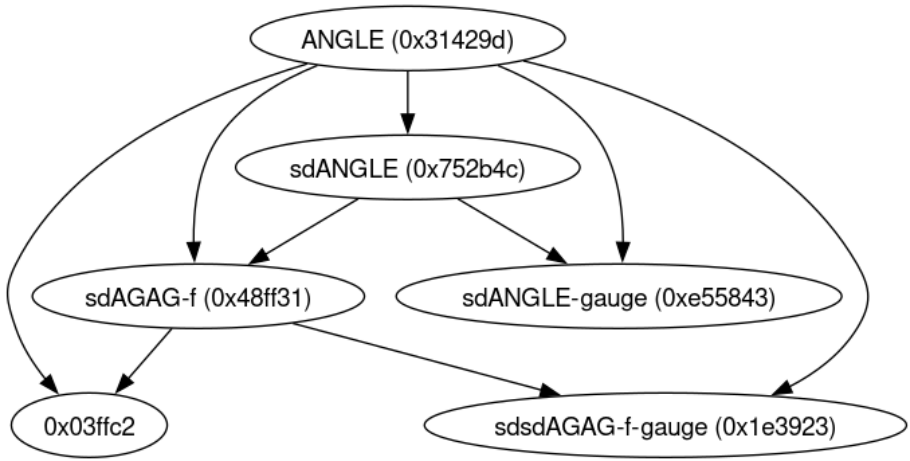

  \centering
  \myincludegraphics[width=\linewidth]{graphviz/defi/angle-protocol}
  \caption{The unfiltered token graph has a weakly connected component
    whose vertices correspond to tokens related to the Angle Protocol
  and Stake \glstext*{dao}\@.}\label{fig:angle-protocol}
\end{figure}

There are \num{418} components, other than the giant component, with
more than two connected vertices.  \Cref{fig:angle-protocol} is an
example from this set.  It shows the tokenising relationships between
tokens related to the Angle Protocol~\citep{angle-xx} and Stake
\glstext*{dao}~\citep{stake-dao-xx}.  Stake \glstext*{dao} implements
investment strategies based on other decentralised protocols.  Their
\textquote{Liquid Lockers} generate liquidity, voting power, and
yield from lockable tokens.  \unit{\angle} is Angle's governance token.  It
can be deposited with Stake \glstext*{dao} to mint \unit{\sdangle}\@.  It
is not possible to burn \unit{\sdangle} and withdraw \unit{\angle}:
the edge between those vertices represents a one-way operation and is
not present in the filtered token graph. \unit{\angle} and
\unit{\sdangle} can be deposited in a gauge (liquidity pool) to mint
\unit{\sdanglegauge}.  We use this visualisation to identify the
direct and transitive dependencies of tokenising tokens such as
\unit{\sdangle} and \unit{\sdanglegauge}.  Many of the other
components in the set produce similar insights relating to other
tokens and protocols.

\begin{figure*}
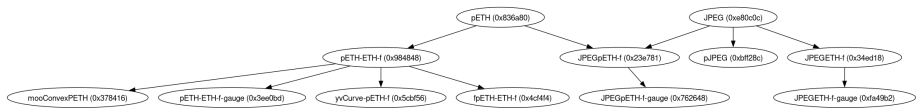

  \centering
  \myincludegraphics[width=\linewidth]{graphviz/defi/jpegd-protocol}
  \caption{The filtered token graph contains a weakly connected
    component whose vertices correspond to tokens related to the
  JPEG'd Protocol.}\label{fig:jpegd-protocol}
\end{figure*}

The filtered graph has \num{1491} weakly connected components.  A
giant component contains \num{4648} vertices (\qty{\sim55}{\percent}) and
\num{5247} edges (\qty{\sim70}{\percent}).  There are \num{1162}
components (\qty{\sim78}{\percent} of the total) with only two connected
vertices.  There are \num{196} components, other than the giant
component, with more than two connected vertices.
\Cref{fig:jpegd-protocol} is an example from this set that relates to
the JPEG'd Protocol~\citep{jpegd-xx}.  JPEG'd is a decentralised
lending market where users can supply \glspl*{nft} as
collateral and obtain loans in \unit{\peth}, a synthetic token that
tracks the price of \unit{\eth}\@.  \unit{\jpeg} is the protocol's
governance token.  The figure shows the various ways in which
\unit{\peth} and \unit{\jpeg} can be tokenised by liquidity pools such
as Curve's \unit[per-mode=symbol]{\peth\per\weth}
pool~\citep{curve-xx}.  Unlike the operations represented by edges in
\cref{fig:angle-protocol}, all operations represented by edges in
\cref{fig:jpegd-protocol} can be reversed by either depositing the
underlying and minting the share, or burning the share and withdrawing
the underlying.

Both the unfiltered and filtered graphs contain giant weakly connected
components.  Exploring such structures requires an interactive user
interface with navigational aids such as selection, panning, and
zooming.  To illustrate the structure in
the giant components, we present the longest directed path in the
filtered token graph.  It comprises nine vertices and represents the
following sequence of tokens:

\begin{enumerate}
  \item \unit{\renbtc}~(\texttt{0xeb4c27})
  \item \unit{\sbtc}~(\texttt{0xfe18be})
  \item \unit{\crvrenwsbtc}~(\texttt{0x075b1b})
  \item \unit[per-mode=symbol]{\tbtclower\per\sbtccrv}~(\texttt{0x64eda5})
  \item \unit[per-mode=symbol]{\btbtc\per\sbtccrv}~(\texttt{0xb9d076})
  \item \unit{\ibbtc}~(\texttt{0xc4e159})
  \item \unit{\wibbtc}~(\texttt{0x8751d4})
  \item \unit[per-mode=symbol]{\ibbtc\per\sbtccrvf}~(\texttt{0xfbdca6})
  \item \unit[per-mode=symbol]{\bibbtc\per\sbtccrvf}~(\texttt{0xae96ff})
\end{enumerate}

Each token in the sequence can be deposited with a contract to mint
the next, and the reverse operations can also be performed (withdraw
\& burn).  The chain involves tokens from several
different protocols.  It is an example of \textquote{composition in
the wild}, groups of interacting smart contracts that span
multiple protocols.

\subsection{Cyclic Structure}\label{sec:analysis-cyclic-structure}

The example token graph in \cref{fig:token-graph} contains an
undirected cycle ($t_0, t_5, t_6$), a directed cycle ($t_1, t_6$), and
a loop ($t_7$).  Both the unfiltered and filtered token graphs contain
undirected cycles.  For example, \cref{fig:angle-protocol} contains
(\unit{\angle}, \unit{\sdangle}, \unit{\sdanglegauge}).  \textit{A
priori}, it
was not apparent if the graphs would contain directed cycles or loops.
The filtered token graph contains neither directed cycles nor loops.
That is, there is no sequence of $n$ tokens represented by vertices
$t_0, t_1, \ldots, t_{n - 1}$ such that $t_i$ is tokenised by $t_{(i +
1) \bmod n}$, $0 \le i < n$.  Is this due to a technical limitation
in our tokenising meta-event heuristic, or is it the case that no such
contracts have been deployed during the time period covered by the
data?  We believe the answer is the latter since it \emph{is} possible
to deploy a contract that creates tokenising meta-events that produce
a directed cycle in the filtered token graph.  We have created such a
contract\footnote{The contract is published at
  \url{https://github.com/harrigan/tokenised-tokens-contracts}: The
  repository contains a smart contract (\texttt{ERC20ExchangeWrapper})
  that can tokenise any token (\texttt{IERC20 underlyingToken}) with any
other (\texttt{IERC20Wrapper overlyingToken}).} to demonstrate this.
It can produce directed cycles of arbitrary length (including loops)
in the filtered token graph.

The unfiltered token graph does not contain any loops.  However, it
does contain a small number of directed cycles.  The graph has
\num{50} non-trivial strongly connected components, that is, strongly
connected components with more than one vertex.  We manually
investigated the tokens involved in these directed cycles.  Many are
\textquote{test tokens} or tokens from defunct protocols, for example,
\unit{\tst}~(\texttt{0x50e508}) and
\unit{\tsttwo}~(\texttt{0x70b34d}), \unit{\tsh}~(\texttt{0x46bada})
and \unit{\tch}~(\texttt{0x2fe3e4}), etc.  All have zero market
capitalisation according to CoinGecko, and only two
(\unit{\usdx}~(\texttt{0x2f6081}) and
\unit{\xbond}~(\texttt{0xa8f8dc})) are traded in liquidity pools
according to \glstext*{dex} Screener.  Although directed cycles and loops can
occur in a token graph, they are not commonplace.

\section{Conclusions}\label{sec:conclusions}

In this paper we present a graph representation of token composition
on the Ethereum blockchain.  We extract tokenising meta-events from
\glstext*{evm} logs, define the token graph, and relate its macro-
and micro-topological structure to the tokenisation process.  The
unfiltered token graph contains \num{23687} tokens and \num{23549}
composition edges; the filtered token graph, which retains only pairs
where both deposit \& mint and withdraw \& burn actions are observed,
contains \num{8424} tokens and \num{7536} edges.  Both have giant
weakly connected components.  Base tokens such as \unit{\usdc},
\unit{\dai}, and \unit{\weth} have high out-degree.  Stablecoins
occupy a distinctive position with high in- and out-degrees, since
they can be minted from various forms of collateral and themselves
used as collateral.

Composition rarely stops at one layer.
\unit{\stkcvxcrvrenwbtcabra}~(\texttt{0xb65ede}), for example,
represents a staked deposit of a share of a liquidity pool for
synthetic and wrapped versions of Bitcoin's native token; its base
tokens include \unit{\renbtc}~(\texttt{0xeb4c27}) and
\unit{\wbtc}~(\texttt{0x2260fa}).  The longest directed path in the
filtered token graph spans nine tokens drawn from several protocols,
an example of \textquote{composition in the wild} where dependencies
traverse organisational boundaries.

Several limitations remain.  Our heuristic for tokenising meta-events
admits false positives, such as gas-token burns and swap-on-deposit
transactions, and false negatives, such as compositions that do not
match the deposit \& mint or withdraw \& burn patterns.  The
CoinGecko and \glstext*{dex} Screener data are snapshots that are not
contemporaneous with the on-chain data.  Our analysis covers Ethereum
mainnet only, while much recent token composition activity occurs on
layer-two networks.

In future work we will refine the heuristic for identifying
tokenising meta-events to reduce the number of false positives, and
extend the analysis to layer-two networks.  The temporal dimension of
the token graph is also unexplored: edges and vertices appear and
disappear over time, and the evolution of the graph could expose the
lifecycles of tokens and protocols.  Finally, whether the filtered
token graph remains acyclic as token composition patterns develop is
an open question.

\bibliographystyle{my-abbrvnat}
\bibliography{tokenised-tokens}

\end{document}